\documentclass{rspublic}

\newcommand{\ds}{\displaystyle}
\newcommand{\reff}[1]{\mbox{\rm (\ref{#1})}}
\newcommand{\mfrac}[2]
{\raisebox{0.045em}{\mbox{\footnotesize$\displaystyle
\frac{#1}{#2}$}}}

\newcommand{\Oint}{\mbox{\footnotesize\raisebox{0.1em}{$\bigcirc$}}
\hspace{-3.9mm}\int}
\newcommand{\sinfty}{\mbox{\tiny$\infty$}}
\newcommand{\pot}{\mbox{\footnotesize $[$} u \mbox{\footnotesize $]$}}

\begin{document}

\title[What integrability means?]
{What does integrability of\\finite-gap or soliton potentials mean?}

\author[Yu. Brezhnev]{Yurii V. Brezhnev\footnote{Permanent address:
\textit{Theoretical Physics Department, Kaliningrad State University,
Kaliningrad 236041, RUSSIA}}}

\affiliation{Department of Mathematics \& Statistics\\
Boston University, Boston MA 02215, USA}

\label{firstpage}

\maketitle

\begin{abstract}{finite-gap integration; algebraic-geometric methods;
soliton theory; algebraic curves; spectral problems;
integrability by quadratures; Liouvillian solutions;
Riccati's equation; trace formulas;
$\Theta$-functions; Abelian integrals}

\noindent In the example of the Schr\"odinger/KdV equation we treat
the theory as equivalence of two concepts of Liouvillian
integrability: quadrature integrability of linear differential
equations with a parameter (spectral problem) and Liouville's
integrability of finite-di\-men\-sional Hamiltonian  systems
(stationary KdV--equations). Three key objects in this field: new
explicit $\Psi$-function, trace formula and the Jacobi problem
provide a complete solution. The $\Theta$-function language is
derivable from these objects and  used for ultimate representation
of a solution to the inversion problem. Relations  with
non-integrable equations are discussed also.

\end{abstract}

\section{Introduction}

As it is understood in contemporary language,
the theory of  algebraic-geometric (or finite-gap)
integration is a theory of integration of
integrable nonlinear (1+1)-soliton partial differential
equations ({\sc pde}'s), finite-dimensional Hamiltonian
dynamical systems,  and
spectral problems defined by ordinary differential equations ({\sc ode}'s).
A large number of papers devoted to these problems, appeared over the last three
decades,
show that the vague term `integration of integrable' does not mean an automatical
integrating procedure in some simple sense of the word.
The contributions that are still being made to this theory  testify to
its vitality: all
the evidence points to the continuance of its growth.

Although the term `completely integrable' is  quite appropriate one,
there is not commonly agreed answer to the questions about
integrability and, in particular, to the question in the title of
the present work. For example, in the theory of dynamical systems
this means the well defined Liouvillian integrability, but the
search for separability variables (followed by action-angle ones) is
a subject of an independent theory (Flaschka \& McLaughlin 1976;
Vanhaecke 1996). In the case of (1+1)-integrable {\sc pde'}s, the
theory is, in fact, a treatment of these equations as
infinite-di\-men\-sional analogues of Hamiltonian systems (Zakharov
\& Faddeev 1971; Gardner 1971) plus set of nontrivial exact
solutions in terms of elementary functions: solitons, positons, and
their relatives (Ablowitz \& Segur 1981). There are natural
multidimensional generalizations of the theory (the
Kadomtsev--Petviashvili ({\sc kp}), Davey--Stewartson equations,
their hierarchies,  etc.) which are rather well developed also.
Integrability of spectral problems (direct and inverse) is usually
associated with the inverse scattering method, soliton theory
(Ablowitz \& Segur 1981; Novikov 1984;  Levitan 1987) and its
$\Theta$-generalizations (Dubrovin 1975; Matveev 1976; McKean \& van
Moerbeke 1975; Flaschka 1975; Novikov 1984; Belokolos {\em et all\/}
1994; Gesztesy \& Holden 2003). We must include here the deep links
of this field to diverse areas of mathematics shedding light on the
mechanism of integrability: Hirota's $\tau$-function method,
Painlev\'e analysis (Conte 1999), Abelian varieties (Dubrovin 1976;
Vanhaecke 1996), Darboux--B\"acklund algebraic transformations
(Matveev \& Salle 1991), differential geometry (Darboux 1915),
theory of commuting differential operators (Krichever 1978) etc.
Each of these links is sufficient in itself to provide a complete
development; combined, they exhibit an unusual wealth of ideas and
furnish rich resources of new interrelations.

Nevertheless, there is a common object for all the approaches
mentioned above: a $\Theta$-representation for a wide class of
solutions and fundamental $\Psi$-function as a solution to
associated linear spectral problem. The key role of this object was
discovered by Its \& Matveev (1975) in the example of the
KdV-equation. The next years after, Krichever (1976, 1977a) put such
a construction into a basis of integration of soliton equations and
it became clear that this link between KdV--equation and the
Schr\"odinger operator
\begin{equation}\label{1}
\Psi''-u\,\Psi=\lambda\,\Psi, \qquad u=\phi(x)
\end{equation}
is not an exception but a common feature.
The ideology is spread to the whole class of such problems.
Presently this is known as a concept of the Baker--Akhiezer function
(Krichever 1978) and such an approach generates all the (1+1)-soliton
equations and their hierarchies if an algebraic curve has been specified.

Originally the term `finite-gap' meant spectral problem for the
smooth real periodic potential with finitely many lacunae at
spectrum of Schr\"odinger's operator. Later, in the 1970's, in works
by Matveev, Its, Dubrovin, Krichever, and others that treatment was
generalized to quasi-periodic complex valued potentials and related
to methods of algebraic geometry and $\Theta$-functions (Matveev
1976; Dubrovin {\em et all\/} 1976; Krichever 1977b). By this reason
throughout the paper we keep traditional terminology `finite-gap'
but identify it with `algebraic-geometric'.

It is difficult to keep pace with the continuing growth of the
literature which is due to the activity of mathematicians. To become
acquainted with the background of this field, as well as find a
complete set of references, it is perhaps best to consult the
surveys written by initiators of the theory.

\section{Solvable potentials for the Schr\"odinger equation}

Until the pioneer work by Novikov (1974) the Schr\"odinger equation
\reff{1} was mostly an object of the spectral theory of operators
(Akhiezer 1961; Levitan \& Sargsjan 1975). By  that time few
solvable (in different senses) examples were known.

\begin{enumerate}

\item The constant potential (trivial case);

\item Quantum harmonic oscillator $u=x^2$;

\item Linear potential $u=x$;

\item Decaying potentials $u=-n\,(n+1)\cosh^{-2}x$
and their rational degenerations;

\item Reflectionless  potentials of Bargmann (solitons) characterized by the
property $\Psi(x;\lambda)=P(x;\lambda)\, \exp\sqrt{\lambda\,}x$ with a
polynomial in $\lambda$ function $P$ (see also Darboux (1915: p.\,212));

\item The time isospectral deformations of these reflectionless potentials
being governed by KdV-dynamics
(Ablowitz \& Segur 1981);

\item Lam\'e's potentials $u=n\,(n+1)\,\wp(x)$
(see  Whittaker \& Watson (1927));

\item Generalizations of the Lame potentials (Darboux 1882, 1915: p.\,228)
$$
u=\mu\,(\mu-1)\,k^2\,\frac{\mbox{cn}^2x}{\mbox{dn}^2x}+
\mu'(\mu'-1)\,k^2\,\mbox{sn}^2x+\frac{\nu\,(\nu-1)}{\mbox{sn}^2x}+
\nu'(\nu'-1)\,\frac{\mbox{dn}^2x}{\mbox{cn}^2x},
$$
being called now  the Treibich--Verdier  potentials. The following
year after Darboux, two comprehensive m\'emoires by Sparre (1883)
appeared on further generalizations.

\end{enumerate}

A considerable result of Matveev and Its  is that they all, apart
from the cases 2--3, lie within the framework of unified theory and
are specifications or degenerations of a wider class, the class of
algebraic-geometric potentials expressible by the following  nice
formula (Its \& Matveev 1975a,b; Matveev 1976)
\begin{equation}\label{theta}
u=-2\,\frac{\rd^2}{\rd
x^2}\ln\Theta\big(x\boldsymbol{U}+\boldsymbol{D}\big)+\const
\end{equation}
The same authors (see also Novikov 1974; Dubrovin 1975; Lax  1975)
proved  that all these potentials are solutions of higher stationary
{\sc ode}'s which are presently named Novikov's equations. An important
observation of Novikov (1974) was that these equations are
representable as Hamiltonian finite-dimensional dynamical systems in $x$.
Soon this  property was completely clarified by Gel'fand \& Dikii (1975, 1979).

The Hamiltonian treatment of Novikov's equations closely joined the
examples mentioned above with the well-known Liouville's integrability.
Moreover,
Liouville's integrability of these equations has received
a natural completion. Namely,
all the solutions are given by the famous trace formula (Matveev 1976)
\begin{equation}\label{trace}
u=2\,\sum\limits_{k=1}^g\,\gamma_k^{}
-\sum\limits_{k=1}^{2g+1} E_k^{}.
\end{equation}
Notice that presence of Liouville's
attributes itself does not automatically provide a procedure of the integration
(this is a theorem of
existence) but the trace formula \reff{trace} supplemented
with the Jacobi  inversion problem brings about such a procedure.
In the language of dynamical systems this means, in fact, transformation to
separability variables.

What could one say about solution  $\Psi$ corresponding to the
potentials mentioned above? The answers are well-known. Elementary
functions in the cases 1 and 4--6, Airy's special functions in the
case 3, the Hermit\'e polynomials (up to an exponential factor) for
the case 2 under $\lambda=-2\,n-1$ with integral $n$, Weber's (or
parabolic's cylinder) special functions  for this case under
arbitrary $\lambda$, and elliptic and related to them functions for
the cases 7--8. For example, in the case 7 with $n=1$ we have
\begin{equation}\label{lame}
\Psi(x;\lambda)=\frac{\sigma\big(x+\wp^{\mbox{\tiny $-1$}}(\lambda)\big)}{\sigma(x)}\re^
{\zeta(\wp^{\mbox{\tiny$-1$}}(\lambda))\,x}.
\end{equation}
The $\Theta$-representation of the $\Psi$-function corresponding to
the potentials \reff{theta}, in traditional notations and
terminology, is given by the formula of A.\,Its:
\begin{equation}\label{Theta}
\Psi\big(x;\lambda(\mathcal{P})\big)=\mathrm{const}(\mathcal{P})\,
\frac{\Theta\big(\boldsymbol{A}(\mathcal{P})+
x\boldsymbol{U}+\boldsymbol{D}\big)}
{\Theta\big(x\boldsymbol{U}+\boldsymbol{D}\big)}\,
\re^{\Omega(\mathcal{P})x}.
\end{equation}

On the other hand, the elementary solutions to the $\Psi$ are representable
by elementary functions, i.\;e. exactly solvable in terms of
elementary indefinite integrals.
The elliptic cases 7--8 and \reff{lame}
are not exceptions: they are representable by indefinite elliptic integrals.
A natural question arises: what about \reff{theta} and \reff{Theta}?
The answer is that these are  not  exception as well. Indefinite quadratures
is a common property of the spectral problem under consideration.
This means that the known Liouville's integrability of nonlinear
Novikov's equations, in fact,
turns out to be equivalent to the quadrature integrability of the
`linear' $\Psi$ and conversely.
In turn, the well-known  effective
solvability of direct/inverse spectral problems in the class of
analytic nonsingular decaying soliton potentials (Ablowitz \& Segur 1981)
turns out to be nothing else but
explicit solvability by \textit{elementary} tools, i.\,e. indefinite
integrations and their inversions. Moreover, such an opportunity is only one.

The next sections contain
proof of the statements above and we suggest that this is a common feature
of all spectral problems arising in the soliton theory.
It should be emphasized here that the representation \reff{Theta}
(after crossing out $\lambda$ and adding the {\sc kp}-variables
$y,\,t_1^{},\,t_2^{},\,\ldots$)
as a function of a point $\mathcal{P}$ on  arbitrary algebraic curve
is a natural and fundamental object for {\sc kp}-hierarchies
of (2+1)-{\sc pde}'s (Krichever 1977).
We shall restrict our consideration only to  spectral problems
defined by {\sc ode}'s as independent objects, so that their
integrability, in the above sense, belongs to their intrinsic nature.
Moreover, we will restrict ourselves only to the Schr\"odinger equation
\reff{1} that we view merely as a differential equation
with a parameter  (subject to explicit integration) rather than a
spectral problem
with any boundary conditions, commuting operator, etc.

\section{Jules Drach}

Perhaps the most surprising facet is the fact that the ideology
mentioned above has not received mention in the modern literature in
the context. It belongs to J.\,Drach (1919) and his name was
revealed by D.\,\& G.\,Chudnovsky (1984) and V.\,Matveev (Belokolos
{\em et all\/} 1994: pp.\;84--85), who drew attention his remarkable
results. The first sentences in (Drach 1919) clearly indicate his
motivations\footnote{\ldots where $h$ is an \textit{arbitrary
parameter}. The most interesting among them are those where the
Riccati equation
$$\rho'+\rho^2=\varphi+h$$ (and consequently the equation
$\frac{d^2y}{dx_{\mathstrut}^2}=[\varphi(x)+h]\,y$ as well) can be integrated by
\textit{quadratures}; we will  show how one determines the
function $\varphi$ in all these cases.}.

\begin{theorem}[Drach's Thesis]
The class of finite-gap $($algebraic-geometric$)$ potentials is the only one
when the spectral problem $(1.1)$ is integrable for the $\Psi$ by quadratures under
all values of the parameter $\lambda$. The solution has the form
\begin{equation}\label{Psi}
\Psi_{\pm}(x;\lambda)=
\sqrt{R(x;\lambda)\,}\,\exp\!\!\int\limits^{\,\,x}
\!\!{\frac{\pm\mu\,\rd x}{R(x;\lambda)}},
\end{equation}
wherein the function $R$ is a polynomial in $\lambda$ solution of
the equation
\begin{equation}\label{curve}
\mu^2=-\frac12\,R\,R''+\frac14{R'}^2+(u+\lambda)\,R^2.
\end{equation}
\end{theorem}
Drach himself does not give a proof therefore  to produce one,
including step-by-step mechanism of integration, is not without
interest. The proof will call to mind the  classical result of Its
\& Matveev (1975) about criteria of the potential to be finite-gap.
However we do not invoke any attributes of spectral theories:
reality of the potential,  periodicity, Weyl's bases, monodromies,
squares of eigenfunctions
$R=a\Psi_1^2+b\Psi_1^{}\Psi_2^{}+c\Psi_2^2$, spectrums, resolvent,
functional spaces, etc.

Actually we will be writing  known and somewhat new formulas of the
theory but keeping in mind only ideology of Drach. Before  passing
to the proof of the theorem we will formulate a known statement
which occurs  in the literature in numerous contexts (Ermakov 1880;
Marchenko 1974;  Its \& Matveev 1975; Gel'fand \& Dikii 1975; Al'ber
1979).

\begin{proposition}
Integrability of the equation \reff{1} is equivalent to integrability
of the third order linear differential equation
\begin{equation}\label{R}
R'''-4\,(u+\lambda)\,R'-2\,u'R=0
\end{equation}
or  compact nonlinear equation of Ermakov for quasi-amplitude
$\Xi=\sqrt{R\,}$ of the $\Psi$$:$
\begin{equation}\label{ermakov}
\Xi''-(u+\lambda)\,\Xi=-\frac{\mu^2}{\Xi^3}.
\end{equation}
\end{proposition}
Though this proposition is a straight consequence of the equations
(\ref{Psi}--2), we will give a {\em derivation\/} all the formulas
(\ref{Psi}--4). Arguments  are as follows.

One sufficient test of quadrature integrability   a given {\sc ode}
is furnished by a solvable Lie point symmetry
of the equation. This technique (group analysis of differential equations)
is rather well developed (Ibragimov 1985; Eisenhart 1933)
so that applying this simple computations to the
Schr\"odinger equation \reff{1}  we get
necessary attributes of the theory in a natural way.
Indeed, generator $\widehat{\boldsymbol{\mathfrak{G}}}$
of the point symmetries $(x,\,\Psi)
\stackrel{\varepsilon}{\mapsto}
(\widetilde x,\,\widetilde\Psi)$
for the equation \reff{1} has the following form:
\begin{equation}\label{G}
\widehat{{\mathfrak{G}}}= \big(a\,\Psi +2\,R\big)\,\partial_x+\big\{
a_x\Psi^2+ (R'+\const)\,\Psi+b\big\}\,\partial_\Psi^{},
\end{equation}
wherein $a=a(x;\lambda),\,b=b(x;\lambda)$ are arbitrary solutions of
\reff{1} and $R=R(x;\lambda)$ satisfies the equation \reff{R}. The
functions $a(x;\lambda),\,b(x;\lambda)$ do not help in further
integrating (need to know solution $\Psi$ itself) and we set them
equal to zero. The remaining free constant in \reff{G} says that the
symmetry does not disappear and we have a solvable commutative
2-parametric symmetry $\big\langle
\widehat{\boldsymbol{\mathfrak{G}}}_1^{}= \Psi\,\partial_\Psi^{},\;
\widehat{\boldsymbol{\mathfrak{G}}}_2^{}=
2R\,\partial_x+R'\Psi\,\partial_\Psi^{}\big\rangle$. Transformation
to `integrable' variables $(x,\,\Psi) \mapsto (z,w)$ is provided by
the standard Lie symmetry machinery. We get an explicit change
(Eisenhart 1933: p.\,91, case $3^\circ$)
$$
z=\int\limits^{\,\,x}\!\!\frac{\rd x}{R(x;\lambda)}, \qquad
w=\ln\frac{\Psi}{\sqrt{R(x;\lambda)}}.
$$
In new variables, as Lie's theory guarantees, the equation will be
easily solvable. Making use of one-fold integrated form of the
equation \reff{R}, i.\,e. the equations \reff{curve} or
\reff{ermakov}, an equation for the function $w=w(z)$ becomes
$w_\mathit{zz}+{w_z}^{\!\!\!2}=\mu^2$, where the constant of
Ermakov--Drach $\mu$ is a constant of integration. The equation is
readily solved indeed and, after back transformations, we arrive at
the formula \reff{Psi}.
Point of departure of Ermakov (1880) was the same---integrability by quadratures.

\section{Integrability of the Schr\"odinger equation: Theorem 3.1}

\subsection{Proof}
\textit{Necessity}. The function $\Psi$ followed by the $R$ is a
function of two variables $x$ and $\lambda$. Our aim is to find out
the dependence of these functions upon both the variables. They are
analytical entire functions of the parameter $\lambda$ (Yakubovich
\& Starzhinskii 1975). This allows one to represent $R$ in a form of
analytical  series
\begin{equation}\label{series}
R(x;\lambda)=R_0(x)+R_1(x)\,\lambda+R_2^{}(x)\,\lambda^2+\cdots
+R_k(x)\,\lambda^k+\cdots\,.
\end{equation}
The equation \reff{R} is linear hence there exists a recurrence relation on
the coefficients $R_k^{}$ derivable from the equation \reff{R}
\begin{equation}\label{K}
R_{k-1}^{}=\widehat{\mathcal{K}}\,R_k^{},\qquad
\widehat{\mathcal{K}} \equiv \frac14\,\partial_{\mathit{xx}}-u+\frac12
\int\limits^{\,\,x}\!\! u_x\ldots\,\rd x\;.
\end{equation}

The next step is to make use of
Ermakov--Drach's equation \reff{curve} wherein
the constant $\mu$ is, at the moment,  {\em arbitrary and independent of\/} $\lambda$.
We must
require integrability under {\em all\/} values $\lambda$. This means
that after
substitution the series \reff{series} into \reff{curve}, coefficients in front
of $\lambda^k$ must be zeroes, independently of each other:
$$
\Big(\mu^2+\mfrac12 R_0^{}R_0^{}\!\!{''}-\mfrac 14 {R_0'}^2-uR_0^2\Big)+
\Big(\mfrac 12(R_0R_1){''}-\mfrac 32 R_0'R_1'
-2\,uR_0^{}R_1^{}-R_0^2\Big)\lambda+\cdots\;.
$$
Hence, the coefficients $R_k(x)$ are determined by  subsequent
integration this infinite system of chained equations.
It is clear that quadrature integrability takes place not for
arbitrary functions $u=\phi(x)$. By which restrictions
are such functions distinguished from all possible ones?

Rewrite the series \reff{series}  as a power series
in $\zeta=\lambda^{-1}$ (Gel'fand \& Dikii 1975):
$$
R(x;\lambda)=\widetilde R_0(x)+\widetilde R_1(x)\,\zeta+
\widetilde R_2(x)\,\zeta^2+\cdots
+\widetilde R_k(x)\,\zeta^k+\cdots\;,
$$
whereupon the recurrence relation \reff{K} acquires the well-known
computable  form:
$$
\widetilde R_k^{}=\widehat{\mathcal{K}}\,\widetilde R_{k-1}^{}:\quad
R(x;\lambda)=1-
\Big(\mfrac{1}{2}^{}\,u-c_1^{}\Big)\zeta-
\Big(\mfrac {1}{8}\,u_{\mathit{xx}}-\mfrac38\,u^2+
\mfrac12\,c_1^{}u-c_2^{}\Big)\zeta^2+\cdots\;.
$$
Thus, for arbitrary $u=\phi(x)$ the $\Psi$ is expressed by the
formula \reff{Psi} with the differential polynomial
$R(\pot;\lambda)$ of infinite order. Moreover, making use of this
recurrence with  subsequent collection in $\zeta$, the equation
\reff{curve} turns into
\begin{equation}\label{curve2}
\ds\zeta\,\mu^2=1+2\,c_1^{}\zeta+\big(2\,c_2^{}+c_1^2\big)\zeta^2+
2\,\big(c_3^{}+c_1^{} c_2^{}\big)\zeta^3_{\ds\mathstrut}+
\cdots+\mbox{second half}\,.
\end{equation}
The dependence on $u$ has gone at infinity: the first half (it is
infinite!) contains only constants $c_k^{}$ and the second one does
differential polynomials of infinite order in $u$. In order to get
conditions of finite order in derivatives $u^{(k)}$ we have the only
possibility---to set $R$ to be a finite polynomial:
\begin{equation}\label{poly}
R(\pot;\lambda)=\lambda^g+R_{g-1}\lambda^{g-1}+\cdots+R_1 \lambda+R_0.
\end{equation}
Accordingly, splitted restrictions on the potential \reff{curve2}
become $g$, now not necessary zero, constants
$I_k^{}=F_k^{}(c_1^{},\,c_2^{},\ldots,c_g)$ plus $g+1$ differential
conditions of finite order
\begin{equation}\label{novikov}
I_k^{}=F_k^{}\big(u,\,u_x,\ldots,\, u^{(2g)};
\,c_1^{},\,c_2^{},\ldots,\,c_g^{}\big), \qquad k=g+1\ldots 2\,g+1.
\end{equation}
Their compatibility  means that such a class of potentials is not
empty and the parameter {\em $\mu$ must depend on the spectral
one\/}. This corresponds to a choice the particular solution $R$
among three linear independent $R_{1,2,3}$ defined by  integration
constants $A_{1,2,3}$ such that the constant $\mu(A_{1,2,3})$ is
algebraically related with $\lambda$: $\mu=\mu(\lambda)$. These
potentials admits such a dependence and the condition \reff{curve2},
or which is the same \reff{curve}, turns into a hyperelliptic
algebraic curve of  finite genus
\begin{equation}\label{curve3}
\begin{array}{l}
\boldsymbol{\mathfrak{S}}:\qquad
\mu^2=\lambda^{2g+1}+I_1\lambda^{2g}+\cdots+I_{2g}^{}\lambda+
I_{2g+1}=\\
\phantom{\boldsymbol{\mathfrak{S}}:\qquad
\mu^2}=(\lambda-E_1)\cdots(\lambda-E_{2g+1}
{}^{\ds\mathstrut}).
\end{array}
\end{equation}

The set of equations \reff{novikov} is 1-fold
integrated Novikov's equations
(stationary KdV-equations) and a half of their integrals
$I_k^{}=F_k^{}(\pot;c)$.
Thus quadrature integrability of the $\Psi$ has became equivalent to
integrability of these equations. If it has been established,
a final answer is given by substitution
$u=\phi(x)$ into the polynomial \reff{poly} and then into the formula \reff{Psi}.
We emphasize that the well-known formula \reff{Psi}
\textit{does not mean any
integrability} without formula for $\phi(x)$. This is ansatz
and such a form of solution to the {\sc ode} \reff{1} does
exist for arbitrary $\phi(x)$.

\textit{Sufficiency}. The sufficiency is a procedure of explicit
integration of the equations \reff{novikov}. It is widely known and
we do not repeat details here (Drach 1919; Dubrovin 1975). The
answer is the formula \reff{trace} plus integral representation for
the functions $\gamma_k^{}(x)$. Nevertheless we note that  being a
remarkable {\em identity\/} in the spectral theories (Levitan 1987),
the trace formula \reff{trace}
turns into a necessary and key object in the quadrature methodology%
\footnote{Independent role of these formulas for  the KdV--equation,
as well as their dynamical relatives, was pointed out by Matveev
(1975). See also his appendix to (Dubrovin {\em et all} 1976).}.
\hfill$\square$

\subsection{Drach--Dubrovin equations and formula for the $\Psi$-function}
\noindent Rather than mere identities, we consider  consequences of
the well-known definition of the polynomial $R$ and new variables
$\gamma_k^{}$, for example $u_{\mathit{xx}}-3\,u^2 \cong
\sum\gamma_j^{}\gamma_k^{}$, as algebraic  transformations from the
variables $\{u,\,u_x,\ldots\}$ to the separability variables
$\{\gamma,\,\mu\}$. Namely, writing the function
\begin{equation}\label{temp}
\begin{array}{l}
\ds
R(\pot;\lambda\big)=
\lambda^g-\Big(\mfrac u2-c_1^{}\Big)\lambda^{g-1}
-\Big(\mfrac{1}{8^{}_{\ds\mathstrut}}\,u_{\mathit{xx}}-\mfrac38\,u^2+
\mfrac12\,c_1^{} u-c_2^{}
\Big)\lambda^{g-2}+\cdots
\end{array}
\end{equation}
in factorized form $\ds R(\pot;\lambda)=
\big(\lambda-\gamma_1^{}(x)\big)\cdots\big(\lambda-\gamma_g(x)\big)$,
we {\em do define\/} the first half of the change of variables
as {\em zeroes\/} of this $R$:
\begin{equation}\label{change}
\big\{u,\,u_x,\ldots, u^{(2g-1)}\big\} \mapsto
\big\{(\gamma_1^{},\,\gamma_2^{},\ldots,\gamma_g),
\quad (\mu_1^{},\,\mu_2^{},\ldots,\mu_g)\big\}.
\end{equation}
More precisely, the first part of the change \reff{change} is as
follows
$$
\left\{\begin{array}{l}
\ds u_{\phantom{\mathit{xx}}}=\phantom{1}2\,
\sum\limits_{k=1}^g\,\gamma_k^{}+2\,c_1^{},\quad 2\,c_1^{}=
-\sum\limits_{k=1}^{2g+1}E_k^{}\\
\ds u_{\mathit{xx}}=
12\,\sum\limits_{k=1}^g\,\gamma_k^2 -
8 \sum\limits_{k,\,j>k}^g\!\gamma_k^{}\gamma_j^{}-
16\,c_1^{}\,\sum\limits_{k=1}^{g^{\mathstrut}}\,\gamma_k^{}
+4\,c_1^2+8\,c_2^{}\\
\ldots\ldots
\end{array}
\right.
$$
and  not closed due to the odd derivatives of $u$. Missing ones and
therefore the second half is extracted by involving the
Drach--Dubrovin differential equations:
\begin{equation}\label{dubrovin}
\frac{\rd \gamma_k^{}}{\rd x}=
\mfrac{-2\,\mu_k^{}}{\ds\prod\limits_{j\ne
k}{(\gamma_k^{}-\gamma_j^{})}},\qquad
\mu_k^2(x)=\big(\gamma_k^{}(x)-E_1^{}\big)\cdots \big(\gamma_k^{}(x)
-E_{2g+1}^{}\big),
\end{equation}
where constants $E_j^{}$ are functions of Novikov's constants
$c_j^{}$ and the integrals \reff{novikov}. We thus get the remaining
part of the complete change \reff{change}:
$$
\left\{
\begin{array}{l}
\ds u_{x\phantom{\mathit{xx}}}=
-4\,\sum\limits_{k=1}^g\mfrac{\mu_k^{}}
{\ds\prod\limits_{j\ne k}{(\gamma_k^{}-\gamma_j^{})}}
=\left(\!2\,\sum\limits_{k=1}^g\,\gamma_k^{}+
2\,c_1^{}\!\right)_{\!x}\\
\ds u_{\mathit{xxx}}=
\;\,16\,\sum\limits_{k=1}^{g\ds\mathstrut}\mfrac{(2\,c_1^{}-3\,\gamma_k^{})}
{\ds\prod\limits_{j\ne k}{(\gamma_k^{}-\gamma_j^{})}}\,\mu_k^{}+
16\sum\limits_{k,\,j>k}^g\!
\left\{\phantom{\frac{{}^{\mathstrut}_{\mathstrut}}{}}\right.
\!\!\!\mfrac{\mu_k^{}\,\gamma_j^{}}
{\ds\prod\limits_{n\ne k}{(\gamma_k^{}-\gamma_n^{})}}+
\mfrac{\mu_j^{}\,\gamma_k^{}}
{\ds\prod\limits_{n\ne j}{(\gamma_j^{}-\gamma_n^{})}}\!\!\!
\left.\phantom{\frac{{}^{\mathstrut}_{\mathstrut}}{}}\right\}\\
\ldots\ldots
\end{array}
\right..
$$
The equations \reff{dubrovin} are readily rewritten into a promised
integral form
which coincides the Jacobi inversion problem for the hyperelliptic
algebraic curve \reff{curve3}
\begin{equation}\label{jacobi}
\mbox{\small$
\left\{
\begin{array}{ccccl}
\ds
\int\limits^{\;(\gamma_1^{},\,\mu_1^{})}\frac{\rd z}{w}\phantom{z}
&\!\!\!\!\!+\;\cdots\;+&
\ds
\!\!\!\!\!\int\limits^{\;(\gamma_g,\,\mu_g)}\frac{\rd z}{w}
&=&d_1^{}\\
\quad\ds{}^{{}^{\mathstrut}}&\!\!\!\!
&&&\\
\ds \int\limits^{\;(\gamma_1^{},\,\mu_1^{})}z\,\frac{\rd
z}{w}\phantom{z} &\!\!\!\!\!+\;\cdots\;+& \ds
\!\!\!\!\!\int\limits^{\;(\gamma_g,\,\mu_g)}z\,\frac{\rd z}{w}
&=&d_2^{}\\
\quad\ds\ldots\ldots{}^{{}^{\mathstrut}}&\!\!\!\!
\ldots&\ldots\ldots&&\ldots\\
\ds \int\limits^{\;(\gamma_1^{\mathstrut},\,\mu_1^{})}
z^{g-1}\,\frac{\rd z}{w} &\!\!\!\!\!+\;\cdots\;+& \!\!\!\!\!\ds
\int\limits^{\;(\gamma_g,\,\mu_g)}z^{g-1}\,\frac{\rd z}{w}
&=&d_g-2\,x
\end{array}
\right.$}.
\end{equation}
With this equation we can obtain formula for the $\Psi$-function
\reff{Psi}  which is not seem to be in the literature:
\begin{equation}\label{summa}
\Psi_{\pm}(x;\lambda)=\exp\frac12\!
\left\{\phantom{\frac{{}^{\mathstrut}_{\mathstrut}}{}} \right.
\!\!\!\!\int\limits^{\;\gamma_1^{}(x)}
\!\!\!\frac{w\pm\mu}{(z-\lambda)\,w}\,\rd z\;+ \cdots+
\int\limits^{\;\gamma_g(x)}
\!\!\!\frac{w\pm\mu}{(z-\lambda)\,w}\,\rd z
\left.\phantom{\frac{{}^{\mathstrut}_{\mathstrut}}{}}
\!\!\!\right\}.
\end{equation}
The variables of integration lie on the curve
$w^2=(z-E_1)\cdots(z-E_{2g+1})$.

It may be remarked at once that complete set of the transformations
above to the separability variables $\{\gamma,\,\mu\}$ is not
necessary for Novikov's equations themselves. This is an attribute
of their Hamiltonian description (Gel'fand \& Dikii 1979) derivable
from the main trace formula \reff{trace} and the equations
\reff{dubrovin}. It thus appears that the  formula \reff{Psi} or
\reff{summa}  supplemented with the two objects \reff{trace} and
(4.9--10) explains the nature of integrability in question.
Subsequent procedure of inversion for a symmetrical sum of
$\gamma$'s is necessary for ultimate representation to the solution.
In the degenerated cases (constant potential, solitons and the like)
all the integrals \reff{jacobi} reduce to integrals of rational
functions, so that the problem becomes trivial (inversion of
logarithms) and leads to well-known exponents. In general case, this
transcendental problem is solved with the help of Riemann's
$\Theta$-functions (see \S\,7).

We would like to note that even `naive'  (not finite-gap) procedure
of integration of the trivial potential
$\Psi''-\const\Psi=\lambda\Psi$ completely accords with the scheme
described above. Holomorphic integrals do not appear but, instead,
the following objects arise: integral definition of the logarithm,
i.\,e. integration of a trivial rational function, and necessity of
inversion of the former (the exponent).

\section{Some consequences}

\subsection{Integrable $\lambda$-pencils}

\noindent
Quadrature integrability does not depend on a choice of
dependent/independent variables. We could formally  avoid the inversion of Abelian
integrals \reff{jacobi} rewriting the theory in the `inverse'
variable $u$. Indeed, after the change $(x,\, \Psi) \leftrightarrow (u,\, Y)$
\begin{equation}\label{5.0}
x=\chi(u),\qquad \Psi=\sqrt{\chi_u^{}}\,\,Y
\end{equation}
we arrive at a second order linear {\sc ode} ($\{\chi,u\}$ is the
standard schwarzian)
\begin{equation}\label{new}
Y_{\mathit{uu}} =Q(u;\lambda)\,Y,\qquad
Q(u;\lambda)=-\frac12\big\{\chi,\,u
\big\}+(u+\lambda)\,\chi_u^2
\end{equation}
which can be of interest in its own right
as a new spectral problem (operator $\lambda$-pencil) for some `good' functions
$Q(u;\lambda)$. The former depends on a chosen  potential $x=\chi(u)$.
Apparently the second equation in \reff{new}, written in the form
\begin{equation}\label{gelfand}
F\,F_{\mathit{uu}}-\frac32\,{F_u}^{\!\!2}-
2\,(u+\lambda)\,F^4+2\,Q(u)\,F^2=0, \qquad F \equiv \chi_u^{}
\end{equation}
and  equation \reff{star} play an important part in the
theory\footnote{See also a footnote on p. 97 in (Gel'fand \& Dikii 1975)
apart from two misprints in one term.}, if only because
inversion of arbitrary finite-gap potential
satisfies this integrable
2-nd order nonlinear {\sc ode} with suitable  function $Q$.
Let us consider some examples $\big(N=n\,(n+1)\big)$.
\begin{itemize}
\item Soliton potentials $u=-n\,(n+1)\,\mbox{cosh}^{-2}x$:
$$
Q(u;\lambda)=-\frac{3}{16}\,\frac{1}{(u+N)^2}
+\frac{\lambda-1}{4\,u^2}
-\frac{\lambda-N-1}{4\,u\,(u+N)};
$$
\item The Lam\'e potentials $u=n\,(n+1)\,\wp(x;g_2^{},g_3^{})$
(Whittaker \& Watson 1927):
$$
Q(u;\lambda)=-\frac{3}{16}\,\sum\limits_{k=1}^{3}\frac{1}{(u-Ne_k^{})^2}+
\frac18\,
\frac{(3+2\,N)\,u+2\,N\,\lambda}{(u-N e_1^{})(u-N e_2^{})(u-N e_3^{})};
$$

\item Arbitrary even elliptic finite-gap potential $u=U\!\big(\wp(x)\big)$, where
$U$ is a rational function of $\wp$. A wide family of such
potentials there provides the theory of elliptic solitons (Acta
Appl. Math. 1994). The function $Q$ is as follows
$$
Q=\frac12\,\frac{\ds \big\{U,\,\wp \big\}}{{U_\wp}^{\!\!\!2}}
+\frac{\ds U\!\big(\wp(x)\big)-3\,\wp(2x)+\lambda}{ {U_\wp}^{\!\!\!2}\,\wp_x^2}=
\widetilde{U}\!\big(\wp(x);\lambda\big),
$$
where $\widetilde{U}$ is another rational function of $\wp$.
Hence $Q=\widetilde{U}\!\big( U^{-1}(u);\lambda\big)$ is
a genus zero algebraic function of $u$
or  rational function of $\wp$: $Q=\widetilde Q(\wp;\lambda)$;

\item Arbitrary elliptic soliton $u=\phi(x)$.
The equation \reff{new} takes the form
$$
Y_{\mathit{uu}}=
\left\{\frac{\Phi_{\mathit{uv}}\,\Phi_u}{v\,{\Phi_v}^{\!\!\!2}}-
\frac12\,\frac{\Phi_{\mathit{uu}}\,{\Phi_v}^{\!\!\!2}+
\Phi_{\mathit{vv}}\,{\Phi_u}^{\!\!\!2}}
{v\,{\Phi_v}^{\!\!\!3}}-\frac14\,\frac{{\Phi_u}^{\!\!\!2}}
{v^2\,{\Phi_v}^{\!\!\!2}}+
\frac{u+\lambda}{v^2} \right\}Y,
$$
where $\Phi(u,v)=0$ denotes a differential equation
connecting the elliptic function $u$ and its derivative $u_x=v$
(algebraic equation of genus unity).
\end{itemize}
By the previous constructions,
all these equations and their relatives of the type  \reff{R}
are of Fuchsian class and  integrable by quadratures. We thus
get a  solution
for all elliptic solitons generalizing formulas of Hermite
(Belokolos {\em et all\/} 1994: p.\,82):
\begin{equation}\label{r}
Y(u;\lambda)=\sqrt{\mathcal{R}(u,v;\lambda)}\,\exp\!\int\limits^{\,u}
\!\!\frac{\mu\,\rd u}{\mathcal{R}(u,v;\lambda)},
\end{equation}
wherein, owing to \reff{temp},
$\mathcal{R}(u,v;\lambda)=v\,R(\pot;\lambda)$ becomes rational function
in $(u,v)$. The last step is a standard problem to
representation the elliptic Abelian integral \reff{r}, which is solved
in terms of Jacobian $\theta$-functions.

It should be remarked
that the presence of explicit $\Psi$ leads to explicit factorization all
of the linear operators \reff{1}, \reff{R} and \reff{new}
with arbitrary $\lambda$'s. For example:
$$
\partial_{\mathit{xx}}-(u+\lambda)=(\partial_x+p)(\partial_x-p)=0,
$$
where roots $\pm p(x;\lambda)$ have the quadrature form
$$
p(x;\lambda)=\frac{\frac12R'+\mu}{R}=
\mu\prod\limits_{k=1}^g(\lambda-\gamma_k^{})^{-1}
-\frac12\,\sum\limits_{k=1}^g
\frac{\gamma_k'}{\lambda-\gamma_k}.
$$
The $\Theta$-functional representation for this and other factorizations
is readily written down using formulas of \S\,7.

\subsection{Liouvillian integrabilities}

\noindent To all appearances Liouville was the first (1833--41) to
recognize a significance of the object $R=\Psi_1^{}\Psi_2^{}$ and
associated linear {\sc ode} of  higher order in the context of
integration of linear {\sc ode}'s in closed form (Liouville 1839).
The third order equation \reff{R} explicitly arose and was discussed
on pp. 430--431 in (Liouville 1839).
Though Liouville was doing in the spirit of algebraic solvability%
\footnote{See however p.\,456 in (Liouville 1839) about what is
nowadays named Liouville's extension.}, a presence of a parameter in
equations turns the theory into the spectral one. The polynomials in
$\lambda$  arose also in works by Darboux and polynomial in $\wp(x)$
was considered by Hermit\'e in the case of Lam\'e's potentials
$u=n\,(n+1)\,\wp(x)$ (Whittaker \& Watson 1927). This corresponds
exactly  to the `polynomial in $u$' cases of Liouville (1839) for
the equation \reff{R}.

V. Kuznetsov (2001, personal communication) pointed out a
relationship of the theory with an algorithm of Kovacic (1986) and,
as we have seen now, this key observation leads to the natural
conclusion:
\begin{itemize}
\item \label{thesis}
{\em Liouvillian integrability of linear {\sc ode}'s {\em (1830--40's)} with a parameter
in finite terms
is equivalent to algebraic integrability by Liouville
{\em (1840--50's)} of nonlinear Hamiltonian
systems}\footnote{We speak here
only about general link
and avoid discussion the rigorous correspondence between finite-gap operators
and Liouvillian solutions, differential algebra (Ritt 1948),
algorithms of Singer (1981),  Kovacic (1986), Picard--Vessiot theory
(van der Put \& Singer 2003), etc.
In particular, we do not touch an important question:
when does  isomorphism between these two Liouvillian integrabilities
take place?
We should mention here some comments about this analogy in
(Morales-Ruiz 1999: pp.\,51--52) (see also van der Put \& Singer 2003).
However main attributes of the theory
(spectral curves, polynomial in $\lambda$, $\Theta$-functions, etc.)
are not discussed in these works.}.
\end{itemize}
The theory and examples above show that this is not a coincidence
and, probably, the modern efficient computer-algorithmic theory,
being applied to both equations (1.1) and \reff{R}, would provide
the independent approaches to generation/classification of
integrable linear operator pencils. For  excellent explanation of
Liouville's ideas see books by Mordukhai-Boltovskoi (1910),  Ritt
(1948), references in works by the authors mentioned in  the last
footnote and the Chapter IX in (L\"utzen 1990). Among other things
this book provides a full account of references for further study.
The next section contains an additional information, examples, and
connections with non-integrable equations.

\section{Integrable cases of Riccati and related equations}

For a simplicity and to avoid lengthening the terminology
we will refer to the second order linear
{\sc ode}'s (or potential) and corresponding to them Riccati's
equations of general form $y_z^{}+a(z)\,y+b(z)\,y^2=c(z)$
as one object.
Well-known transformations between them have a quadrature characterization.

\subsection{Riccati's equations}
\noindent The $Q$-functions corresponding to integrable Riccati's
equations \reff{new} can be rational/algebraic, elementary, or
transcendental. It is rather evident that non-integrable equations
contain integrable subcases. Say,  the quantum harmonic oscillator
$Y_{\mathit{uu}}=\big(u^2-2\,\mu -1\big) Y$ shows that this
potential, seemingly having  nothing in common with the finite-gap
ones, is integrable by quadratures if $\mu$  is an integer.

Let us reverse a view on the equations and change (\ref{5.0}--2).
Whether exists transformation between  nonfinite-gap equation
(with a parameter or no)
\begin{equation}\label{riccati}
Y_{\mathit{zz}}=Q(z)\,Y
\end{equation}
and a finite-gap one? Such a transformation   depends on
chosen equations and can be rather complicated. In contrast to the preceding
(\ref{5.0}--2),
corresponding functional relation is given by the general change of variables
$(x,\Psi)\mapsto(z,Y)$
\begin{equation}\label{nnew}
x=\chi(z),\qquad \Psi=\sqrt{\chi_z^{}}\,\,Y
\end{equation}
and depends on the potential $u=\phi(\chi)$.

\begin{proposition}
Arbitrary  equation \reff{riccati} and the finite-gap one $(1.1)$
are transformable into each other by the following functional relation
\begin{equation}\label{ratio}
\chi:\quad\frac{\Psi_1^{}(x;\lambda)}{\Psi_2^{}(x;\lambda)}=
\frac{Y_1^{}(z)}{Y_2^{}(z)},
\end{equation}
where functions $\Psi_{1,2}^{}(x;\lambda),\,Y_{1,2}^{}(z)$ are
independent solutions of $(1.1)$ and \reff{riccati}.
\end{proposition}
\begin{proof} From (1.1) and (\ref{riccati}--2) we have
\begin{equation}\label{star}
-\frac12\big\{\chi,\,z
\big\}+\big(\phi(\chi)+\lambda\big)\,\chi_z^2=Q(z).
\end{equation}
Clearly, the sought for  functional relation is an integral of this equation.
The potential $\phi(\chi)$ is defined by the corresponding $\Psi$ which is known.
From the second equality in \reff{nnew} we have
$$
\frac{\rd\chi}{\Psi^2}=\frac{\rd z}{Y^2}.
$$
Integrating and  supplementing with the property
$$
\int\!\frac{\rd x}{\Psi_1^2}=\frac{\Psi_2^{}}{\Psi_1^{}}
$$
we get the formula \reff{ratio} and complete the proof.
\end{proof}
It thus appears that the product solution $R=\Psi_1^{}\Psi_2^{}$ is
a fundamental object in the finite-gap theory and  the ratio
\reff{ratio} is fundamental in transformations between Riccati's
equations. Such arguments might  seem to be trivial because every
integrable equation is transformable to the trivial
$Y_{\mathit{zz}}=0$. But an example of Rawson (1883) shows
nontrivial consequences: generating of finite-gap spectral problem
\reff{1} with $\phi(x)=n\,(n+1)\,x^{-2}$ from the oldest and
classical equation of Riccati--Bernoulli with a parameter. The paper
of Rawson is so short that we completely reproduced it in the
Appendix A without any comments\footnote{ See also Liouville's
(1841: pp.\,11--13) classical considerations on integrability of the
potential $\phi(x)=B\,x^{-2}$ with appearance $B=n(n+1)$. Another
example of `triviality' is the class of potentials in elementary
functions (solitons and the like) generated from  the zero potential
by the Darboux transformation (Matveev \& Salle 1991).}.

The following extra examples exhibit a functional
relation between  the Lam\'e potentials
and equidistance spectrum of harmonic oscillator
$Y_{\mathit{zz}}=\big(z^2-2\,\mu-1\big) Y$.
\begin{itemize}
\item $\phi(x)=n\,(n+1)\,x^{-2}$ and $\mu=0$.
We obtain
$$
\chi:\quad x^{2n+1}=\int\limits^{\,z}\!\re^{z^2}\rd z,\qquad \lambda=0;
$$
\item The Lam\'e 1-gap potential $\phi(x)=2\,\wp(x)$.
The  relation \reff{ratio} has the form
$$
\chi:\quad\frac{\sigma(\alpha-x)}{\sigma(\alpha+x)}\,\re^{2\zeta(\alpha)x}=
\int\limits^{\,z}\!\!\frac{\re^{z^2}}{H_\mu^2(z)}\,\rd z,
\qquad \alpha=\wp^{-1}\hspace{-0.2mm}(\lambda).
$$
\end{itemize}
As we have seen, these formulas are representable by quadratures in
both variables. List of examples along these lines may be readily
extended. For example one to derive a generalization of Rawson's
transformation (see  Appendix A) from the classical equation of
Riccati to the Lame one. A counterexample of Liouville--Airy
$Y_{\mathit{zz}}=(z+\alpha) Y$ brings out  the differential field
independently of the parameter $\alpha$:
$$
\chi:\quad
\exp\!\!\int\limits^{\,x}\!\!\frac{2\,\mu\,\rd x}{R(\pot;\lambda)}=
\int\limits^{\,z}\!\!\frac{\rd z}{\mbox{Ai}^2(z+\alpha)},
$$
as might be expected for the variable $z$.

In a broad sense, the variables $u=\phi(x)$ and $\Psi(x;\lambda)$
may be thought of as `convenient' variables for all integrable
Riccati's equations with a nontrivial parameter because, in this
case, the differential polynomial $R(\pot;\lambda)$ has a universal
description as $u=\phi(x)$ is a solution of the equations of
Novikov.

\subsection{Related equations}
\noindent
The equations of Riccati,  Ermakov--Drach, Novikov's
equations, and equations of the type (\ref{5.0}--3) and \reff{star}
are hidden forms of integrability of one another but are not only
integrable linear/nonlinear equations arising in this theory.
Besides them, the $\Psi$-function itself satisfies some nonlinear
homogeneous autonomous differential equation of the third order with
a parameter(s). That equation and its relatives are obtained with
the help of suitable elimination of the potential $u$. The following
instances illustrate the remark above.
\begin{itemize}
\item  $\phi(x)=n\,(n+1)\,\wp(x+c)$:
$$
n\,(n+1)\big(\Psi\,\Psi'''-\Psi'\Psi''\big)^2=
4\,\Psi\big(\Psi''-\lambda_1^{}\Psi\big)
\big(\Psi''-\lambda_2^{}\Psi\big)\big(\Psi''-\lambda_3^{}\Psi\big),
$$
where arbitrary parameters $\lambda_k^{}$ are restricted by the
relation $\lambda_1^{}+\lambda_2^{}+\lambda_3^{}=0$.
\item Arbitrary elliptic finite-gap potential $u$. Then the $\Psi$
satisfies the equation
$$
\Phi\!\left(
\mfrac{\Psi''}{\Psi}-\lambda,\,
\mfrac{\Psi'''}{\Psi}-\mfrac{\Psi'\Psi''}{\Psi^2}\right)
=0,
$$
where $\Phi(u,v)=0$ is an algebraic relation between
$u$ and its derivative $u_x=v$.
\end{itemize}
The result of elimination depends on a chosen independent variable
$x,\,u,\ldots\;$ and all these integrable equations and their
$t$-deformations can be of interest in their own right if only
because they are closely  related to  known 3-rd order autonomous
nonlinear {\sc ode} of Jacobi for the $\vartheta$-constants in the
framework of Fuchsian equations. We do not develop this topic here.
It would appear reasonable  that the widely known and universal
$\Theta$-description of the theory (Krichever 1977a,b) should be
obtainable from the spectral problem itself. This is so indeed.

\section{$\Theta$}

Here we will obtain the $\Theta$-representation \reff{Theta} for the
$\Psi$-function and, thereby, its  properties as a function of
Baker--Akhiezer. Insomuch as regular derivation of this axiomatic
representation is not described in the literature, one to write up
that procedure is, perhaps, not without interest.

As we mentioned above, pure spectral approaches were extended to the
complex valued potentials. (See  recent monograph by Gesztesy \&
Holden (2003) for most exhaustive bibliography and new results in
spectral treatment of the theory). Taking this into account we will
refer a general $\lambda$-dependence of the $\Psi$ as its spectral
property and its dependence upon $x$-variable is considered to be
parametric. Such a spectral view arose in the paper by Akhiezer
(1961) and was completed in full by Its \& Matveev (1975: \S4). It
seems helpful to compare this classical approach with a `reverse'
one, i.\,e. primary $x$-dependence of the $\Psi$. This would
correspond to pure quadrature arguments with a parametrical
$\lambda$-dependence made out in the previous sections.

\begin{theorem}
The $\Theta$-functional representation \reff{Theta} to the
$\Psi$-function is a consequence of the quadrature representations
\reff{Psi} and \reff{summa}.
\end{theorem}

\begin{proof}
Since the spectral parameter
$\lambda$ is connected with the variable $\mu$ by the algebraic equation
\reff{curve3} of finite genus $g$ we view both these variables as
meromorphic functions $\lambda=\lambda(\tau),\,\mu=\mu(\tau)$
of a global parameter $\tau$ on the curve \reff{curve3}.
Accordingly, we consider the
$\Psi$-function \reff{summa} as a function of $x$ and $\tau$
\begin{equation}\label{summanew}
\Psi(x;\tau)=\mbox{\large$\re$}_{}
^{\mfrac12\ds\sum\limits_{k=1}^{g}\!
{\ds \int\limits_{\alpha_k^{}}^{\gamma_k^{}(x)}}
\!\!\!\mfrac{w+\mu(\tau)}{\big(z-\lambda(\tau)\big)w}\,\mbox{\small $\rd z$}},
\end{equation}
where $\alpha_k^{}$ are arbitrary constants. It is a symmetrical
function of the quantities $\gamma_k$ and the formers, as functions
of $x$, are defined from the inversion problem \reff{jacobi}:
\begin{equation}\label{omega}
\mbox{\small$\ds
\left\{
\begin{array}{l}
\ds\int\limits^{\gamma_1^{}}_{\alpha_1^{}}\!\rd \omega_1^{}(z)+\cdots
+\int\limits^{\gamma_g}_{\alpha_g}\!\rd \omega_1^{}(z)=d_1^{}\\
\ds\mathstrut^{\mathstrut}_{\ds\mathstrut}
\ldots\ldots\ldots\ldots\ldots\ldots\ldots\ldots\ldots\ldots\\
\ds\int\limits^{\gamma_1^{}}_{\alpha_1^{}}\!\rd \omega_g(z)+\cdots
+\int\limits^{\gamma_g}_{\alpha_g}\!\rd \omega_g(z)=d_g-2\,x.
\end{array}
\right.$}
\end{equation}
Let $A_{jk}$ be a matrix of $\boldsymbol{a}$-periods
of the holomorphic Abelian integrals \reff{omega}:
$$
\Oint\limits_{a_k^{}}\rd\omega_j^{}(z)=A_{jk}.
$$
All the terminology and notation in this section is standard and  elucidated
in any paper on the finite-gap integration.
As  usual, we normalize the integrals $\omega_j^{}(z)$, introducing the
canonical base of normal holomorphic Abelian integrals
$\widetilde{\boldsymbol{\omega}}(z)$:
\begin{equation}\label{B}
\omega_j^{}(z)=A_{jk}\,\widetilde\omega_k^{}(z), \qquad
\Oint\limits_{a_k^{}}\rd\widetilde\omega_j^{}(z)=\delta_{jk}^{},\qquad
\Oint\limits_{b_k^{}}\rd\widetilde\omega_j^{}(z)=B_{jk}.
\end{equation}
Jacobi's inversion problem \reff{jacobi}, \reff{omega} acquires the
form
\begin{equation}\label{canonical}
\sum\limits_{k=1}^{g}
\widetilde\omega_j^{\gamma_k^{},\alpha_k^{}}=-x\,U_j+C_j,
\qquad U_j=2\big(A^{-1}\big){}_{jg}^{},
\end{equation}
where we adopt the concise Baker's notation (Baker 1897)
for Abelian integrals:
$$
\widetilde\omega_j^{\gamma_k^{},\alpha_k^{}}\equiv
\int\limits^{\,\gamma_k^{}}_{\alpha_k^{}}\!\rd\widetilde\omega_j^{}(z)=
\widetilde\omega_j^{}(\gamma_k^{})-\widetilde\omega_j^{}(\alpha_k^{}).
$$
The vector $\boldsymbol{U}$ depends on the curve
$\boldsymbol{\mathfrak{S}}$ but $\alpha_k^{}$ and $C_j$ are
arbitrary constants. The function \reff{summanew} depends on the
point $\tau$ on the curve $\boldsymbol{\mathfrak{S}}$ and on the
variable $x$ through $\gamma$'s. Thus it is a single-valued function
(but not Abelian) of a point
$\boldsymbol{\eta}(x)=-x\boldsymbol{U}+\boldsymbol{C}$ on the
Jacobian $\mathit{Jac}(\boldsymbol{\mathfrak{S}})$ and, hence, has a
$\Theta$-functional representation. Indeed, the arisen sum
\reff{summanew} is nothing else but the fundamental $T$-function of
Weierstrass\ (1856) and Clebsch--Gordan (Clebsch \& Gordan 1866)
$$
T_{\eta\xi}\!
\binom{\gamma_1^{},\,\gamma_2^{},\,\ldots,\,\gamma_g}
{\alpha_1^{},\alpha_2^{},\,\ldots,\,\alpha_g}=
\int\limits_{\alpha_1^{}}^{\,\gamma_1^{}}\!\rd \widetilde\Pi_{\xi\eta}(z)+
\int\limits_{\alpha_2^{}}^{\,\gamma_2^{}}\!\rd \widetilde\Pi_{\xi\eta}(z)+\cdots+
\int\limits_{\alpha_g}^{\,\gamma_g}\!\rd \widetilde\Pi_{\xi\eta}(z),
$$
where $\rd\widetilde\Pi_{\xi\eta}^{}(z)$ denotes the elementary  normal Abelian
differential of 3-rd kind with 1-st order poles at the points
$z=\xi,\,\eta$ and residues $+1,\, -1$ respectively. Clebsch and Gordan (1866)
devoted Chs.\,6--8 in their book to detail properties of this object
and regular
procedure of derivation the $\Theta$-representation for it
(see also Baker 1897).
We have from there (Clebsch \& Gordan 1866: \S\S 54, 57; Baker 1897:
\S\S 171, 187--188)
$$
\widetilde\Pi_{\alpha,\nu}^{x,z}=
\widetilde\Pi_{\alpha,\nu}(x)-\widetilde\Pi_{\alpha,\nu}(z)=
\ln\!\left[
\frac{\Theta(\widetilde{\boldsymbol{\omega}}^{x,\alpha}+\boldsymbol{r})}
{\Theta(\widetilde{\boldsymbol{\omega}}^{x,\nu}+\boldsymbol{r})}\right.
\!\left/\!
\left.\frac{\Theta(\widetilde{\boldsymbol{\omega}}^{z,\alpha}+\boldsymbol{r})}
{\Theta(\widetilde{\boldsymbol{\omega}}^{z,\nu}+\boldsymbol{r})}
\right]\right.,
$$
where vector
$\boldsymbol{r}=\widetilde{\boldsymbol{\omega}}^{m_g,\,m}-
\widetilde{\boldsymbol{\omega}}^{z_1^{},\,m_1^{}}-\cdots-
\widetilde{\boldsymbol{\omega}}^{z_{g-1},\,m_{g-1}}$ is a zero of
the $\Theta$-function. In our situation we have
$\alpha=\lambda,\,\nu=\infty,\,x=\gamma_k^{},\, z=\alpha_k^{}$. The
representations for zeroes $\boldsymbol{r}$ and the Riemann
constants $\boldsymbol{\mathcal{K}}$ are not unique because they
depend on  a lower bound of the holomorphic integrals
$\widetilde{\boldsymbol{\omega}}(z)$. On the other hand the point
$\boldsymbol{\eta}$ on the Jacobian contains free constants $C_j$ in
\reff{canonical}, so that we may simplify the considerations putting
$m$'s equal to $\gamma_j^{}$'s apart from $\gamma_k^{}$ and set
$$
r_j^{}=\widetilde{\omega}_j^{}(\mathcal{P}_1^{})+\cdots+\widetilde{\omega}_j^{}
(\mathcal{P}_{g-1}^{})+\mathcal{K}_j^{}
$$
with arbitrary points $\mathcal{P}_j$, say $\gamma$'s. We thus
arrived at Riemann's function
$\Theta\big(\widetilde{\boldsymbol{\omega}}(\mathcal{P})-\boldsymbol{e}\big)$.
As a consequence of these arguments and explicit zeroes of the
$\Psi$ in \reff{Psi} we get the following formula (see also Baker
(1928): pp.\,588--589)
$$
\widetilde\Pi_{\lambda,\,\,\infty}^{\gamma_1^{},\alpha_1^{}}+\cdots+
\widetilde\Pi_{\lambda,\,\,\infty}^{\gamma_g^{},\alpha_g^{}}
=
\ln
\frac{\Theta\big(\widetilde{\boldsymbol{\omega}}(\gamma_1^{}) +\cdots+
\widetilde{\boldsymbol{\omega}}(\gamma_g)-
\widetilde{\boldsymbol{\omega}}(\lambda)+\boldsymbol{\mathcal{K}}\big)}
{\Theta\big(\widetilde{\boldsymbol{\omega}}(\gamma_1^{}) +\cdots+
\widetilde{\boldsymbol{\omega}}(\gamma_g)
-\widetilde{\boldsymbol{\omega}}(\infty)
+\boldsymbol{\mathcal{K}}\big)}
+c(x;\tau).
$$
The integrals in \reff{summanew} are elementary but not normal.
Hence we have an identity
$$
\frac12\ds\sum\limits_{k=1}^{g}\! {\ds
\int\limits_{\alpha_k^{}}^{\gamma_k^{}(x)}}
\!\!\!\frac{w+\mu(\tau)}{\big(z-\lambda(\tau)\big)w}\,\rd z=
\widetilde\Pi_{\lambda(\tau),\;\infty}^{\gamma_1^{}(x),\alpha_1^{}}+\cdots+
\widetilde\Pi_{\lambda(\tau),\;\infty}^{\gamma_g(x),\alpha_g}+
\ds\sum\limits_{j,k=1}^{g}\! h_{jk}^{}(\tau)\,
\widetilde{\omega}_j^{}\big(\gamma_k^{}(x)\big)
$$
with some normalizing constants $h_{jk}^{}(\tau)$. Taking into
account Jacobi's problem \reff{canonical} we arrive at the
intermediate answer
\begin{equation}\label{inter}
\Psi(x;\tau)=
\frac{\Theta\big(\widetilde{\boldsymbol{\omega}}(\lambda(\tau))+
x\boldsymbol{U}-\boldsymbol{C}-\boldsymbol{\mathcal{K}}\big)}
{\Theta\big(
\widetilde{\boldsymbol{\omega}}(\infty)+x\boldsymbol{U}-\boldsymbol{C}-
\boldsymbol{\mathcal{K}}\big)}\,\,
\mathrm{exp}\!\!\ds\sum\limits_{j,k=1}^{g}\! h_{jk}^{}(\tau)\,
\widetilde{\omega}_j^{}\big(\gamma_k^{}(x)\big)\,.
\end{equation}

Meromorphic part of the $\Psi$-function has been determined. In
order to determine the function $ f(x;\tau)=\ds\sum
h_{jk}^{}(\tau)\, \widetilde{\omega}_j^{}\big(\gamma_k^{}(x)\big) $
in \reff{inter} we involve the transformation properties of the
$\Psi$ as the function \reff{summanew} on the curve
$\boldsymbol{\mathfrak{S}}$. Let $\widehat a_k^{}\tau$ and $\widehat
b_k^{}\tau$ denote linear-fractional transformations of the variable
$\tau$ corresponding to the cycles $a_k^{}$ and $b_k^{}$. From the
primary representation \reff{Psi} we conclude that $\Psi(x;\tau)$
saves its own form \reff{Psi}, up to a multiplier $M(\tau)$,
 when $\tau$ undergoes the
transformations $\widehat {\boldsymbol{a}},\,\widehat
{\boldsymbol{b}}$. Hence the formula \reff{inter} must hold this
property. Clearly, the function $f(x;\tau)$ is an entire function of
$x$ since $\widetilde{\boldsymbol{\omega}}(\gamma)$'s are everywhere
finite. Further, the $\widehat b_k^{}$-transformations of the
$\Theta$'s in \reff{inter} say that the function $f$ has to be a
linear function in $x$ to compensate an exponential multiplier
$c(\tau)\exp(-2\,\pi\,\ri\, x\,U_k)$ in \reff{inter}:
$$
f(x;\tau)=\varkappa(\tau)\,x+\mathrm{const}(\tau)\,.
$$
Invariance of
the $\Theta$-functions in \reff{inter}  with respect to $\widehat
{\boldsymbol{a}}$-transformations implies
$$
1): \qquad\varkappa(\widehat a_k^{}\,\tau)=\varkappa(\tau)\,.
$$
The transformations $\widehat{\boldsymbol{b}}$ imply
$$
2): \qquad\varkappa\big(\widehat
b_k^{}\,\tau\big)=\varkappa(\tau)+2\,\pi\,\ri\,U_k^{}\,.
$$
Let $\tau$ approaches  the pole $\tau_{\sinfty}$ of the meromorphic
function $\lambda(\tau)$. Such a pole is only one and, designating
$\xi\equiv\tau-\tau_{\sinfty}$, we have
$$
\lambda(\tau)=\frac{A^2}{\xi^2}+
\frac{B}{\xi}+C+\cdots,
\quad
\pm\mu(\tau)=\left(\frac{A}{\xi}\right)^{\!2g+1}+
\frac{2\,g+1}{2}\,\frac{B}{A}\left(\frac{A}{\xi}\right)^{\!2g}+\cdots\;.
$$
From these formulas and \reff{Psi} we obtain (extracting terms independent of
$\gamma$'s)
$$\mbox{\small$
\begin{array}{l}
\ds
\ln\!\Psi(x;\tau)=
\frac12\,\sum\limits_{k=1}^{g}\ln\!\left\{
\frac{A}{\xi^2}+\frac{B}{\xi}+(C-\gamma_k^{})+\cdots\right\}+
\int\limits^{\,\,x}
\!\!\frac{\pm\mu\,\rd x}{(\lambda-\gamma_1^{})\cdots
(\lambda-\gamma_g^{})}=\\
\ds\phantom{\ln\Psi(x;\tau)}= \ds\mathrm{const}(\tau) -
\left\{\frac{\sum\gamma_k^{}}{2A^2}\,\xi^2+
\cdots\right\}^{\mathstrut}\pm
\int\limits^{\,\,x^{\mathstrut}}\!
\left\{\frac{\sum \gamma_k^{}}{A}\,\xi+\cdots \right\}\!\rd x
\ds
\pm\left\{\frac{A}{\xi} +\frac{B}{2A}+\cdots
\right\}^{\ds\mathstrut}x
\end{array}$}
$$
and therefore we get one more property
\begin{equation}\label{mero}
3):\qquad \pm\varkappa(\tau)=\frac{A}{\tau-\tau_{\sinfty}}+\frac{B}{2A}+
\cdots\;.
\end{equation}

A function with the properties 1--3) does exist on $\boldsymbol{\mathfrak{S}}$.
This is the
normal elementary Abelian integral of the second kind
$\pm\varkappa(\tau)=\widetilde\Omega(\tau)$ with the first order pole
at the point
$\tau_{\sinfty}$. Its principal part and a constant in \reff{mero} are
well defined
so that this function is completely determined and unique.
In turn, the three parameters
$\{\tau_{\sinfty},\,A,\,B\}$ may be freely chosen, say, $\{0,\,1,\,0\}$
respectively. The vector $\boldsymbol{U}$ in (\ref{canonical}--5)
becomes the vector of
$\boldsymbol{b}$-periods of the integral $\widetilde\Omega(\tau)$.
Summarizing the arguments above and recovering a normalizing constant
\begin{equation}\label{PsiTheta}
\Psi\big(x;\lambda(\tau)\big)=
\frac{\Theta\big(\widetilde{\boldsymbol{\omega}}(\infty)
-\boldsymbol{D}\big)}
{\Theta\big(\widetilde{\boldsymbol{\omega}}(\lambda(\tau))-\boldsymbol{D}\big)}
\,
\frac{\Theta\big(\widetilde{\boldsymbol{\omega}}(\lambda(\tau))+
x\boldsymbol{U}-\boldsymbol{D}\big)}
{\Theta\big(\widetilde{\boldsymbol{\omega}}(\infty)
+x\boldsymbol{U}-\boldsymbol{D}
\big)}\,
\re^{\widetilde\Omega(\tau)\hspace{0.3mm}x}
\end{equation}
we get the spectral properties of the $\Psi$ and complete the proof.
\end{proof}
From the last expansion of the $\ln\!\Psi(x;\tau)$ we  obtain the
formula \reff{theta}. Indeed,
$$
\int\limits^{\,x}\sum\limits_{k=1}^{g}
\gamma_k^{}\,\rd x=\frac{\rd}{\rd\xi}\Big\{\!\ln
\!\Psi(x;\tau)-\widetilde\Omega(\tau)\,x\Big\}\Big|_{\xi=0}
=
\left.\frac{\rd}{\rd\xi}
\ln\!\frac{\Theta\big(\widetilde{\boldsymbol{\omega}}(\lambda(\tau))+
x\boldsymbol{U}-\boldsymbol{D}\big)}
{\Theta\big(\widetilde{\boldsymbol{\omega}}(\lambda(\tau))-\boldsymbol{D}\big)}
\right|_{\xi=0}
$$
hence, by virtue of the property $\boldsymbol{U}=
-\mfrac{\rd\widetilde{{\boldsymbol{\omega}}}(\tau)}{\rd\tau}\Big|_
{\tau=\tau_{\sinfty}}$, the derivative
$\mfrac{\rd}{\rd\xi}\Big|_{\xi=0}$ may be replaced by $-\rd/\rd x$
and we get the final formula
\begin{equation}\label{Trace}
u=-2\,\frac{\rd^2}{\rd x^2}
\ln\Theta\big(x\boldsymbol{U}
+\widetilde{\boldsymbol{\omega}}(\infty)-\boldsymbol{D}\big)-
\sum\limits_{k=1}^{2g+1}E_k^{}.
\end{equation}

We should conclude here that  the spectral and quadrature
considerations are mutually replaceable. The explicit transition
between these approaches there provides the known Weierstrass's
theorem on a permutation of arguments and parameters in the normal
Abelian integrals of  3-rd kind
$$
\widetilde \Pi_{\nu,\mu}(z)- \widetilde \Pi_{\nu,\mu}(x)=
\widetilde \Pi_{z,x}(\nu)- \widetilde\Pi_{z,x}(\mu).
$$
This fact immediately leads to the equivalence of the `spectral'
formula for the $\Psi$ (Its \& Matveev 1975: formulas (4.12--13))
and the quadrature one  \reff{summa} or \reff{summanew}. The above mentioned theorem, in our notation, has the form
$$
\frac12\!\!\int\limits_{\alpha_k^{}}^{\;\gamma_k^{}(x)}
\!\!\!\!\frac{w+\mu}{(z-\lambda)}\frac{\rd z}{w}=
\frac12\,\int\limits_{\infty}^{\,\lambda}
\!\left\{{}_{\ds\mathstrut}^{\ds\mathstrut}\right.\!\!\!
\frac{w+\mu_k^{}(x)}{\big(z-\gamma_k^{}(x)\big)}-
\frac{w+\beta_k^{}}
{(z-\alpha_k^{})}\!\left\}{}_{\ds\mathstrut}^{\ds\mathstrut}\right.
\!\!\frac{\rd z}{w}+
\mbox{holomorphic part}(x),
$$
where $\beta_k^2=(\alpha_k^{}-E_1)\cdots (\alpha_k^{}-E_{2g+1})$.
Since the $\Psi$ contains all the information, we can obtain
suitable expressions for all objects of the theory. In particular,
$\Theta$-functional representation for `finite-gap function' of
Ermakov \reff{ermakov} having numerous applications. Renormalizing
it by the formula $\Xi^2=\Psi_+(x;\lambda)\, \Psi_-(x;\lambda)$ we
obtain
$$
\Xi^2(x;\lambda)=\const\,
\frac{\Theta\big(\widetilde{\boldsymbol{\omega}}(\lambda)+
x\boldsymbol{U}-\boldsymbol{D}\big)\,
\Theta\big(\widetilde{\boldsymbol{\omega}}(\lambda)-
x\boldsymbol{U}+\boldsymbol{D}\big)}
{\Theta^2\big(\widetilde{\boldsymbol{\omega}}(\infty)
+x\boldsymbol{U}-\boldsymbol{D} \big)}\! .
$$
`Finite-gap' means that the constant $\mu$ in Ermakov's equation
\reff{ermakov} is not independent of $\lambda$: $\mu=\mu(\lambda)$.
Note that the equation \reff{R} itself exemplifies the integrable
and factorizable linear operator pencil  with Abelian coefficients
(like soliton spectral problems) but its solution, as a
counterexample, is not a function of Baker--Akhiezer.

Many of constructions in \S\S\,3, 5--7 can be carried over to
arbitrary spectral problems although not so simply as in the case of
the Schr\"odinger/KdV equation. Nontrivial examples of the $\Psi$
were obtained in (Ustinov 2002) and modifications of Dubrovin's
equations and  trace formulas in (Brezhnev 2002).

\section{Conclusive comments and bibliographical remarks}

Apparently Baker (1928: p.\,587) was the first to realize the
exponential property of the fundamental $T$-function in disguise,
however explicit meromorphic integrals like $\widetilde\Omega(\tau)$
are, to all appearances, the result of the modern theory and became
a universal property of all the integrable models (Krichever
1970's). Note, that throughout the paper we made no restrictions on
the curve \reff{curve3}. It may be singular and the theory can be
rewritten (with minor changes in \S\,7) because the degenerated
holomorphic integrals $\widetilde{\boldsymbol{\omega}}$, in this
case, turn into the integrals $\widetilde{\Pi}$ of the 3-rd kind.
Clebsch \& Gordan (1866: \S43) call the corresponding problem
`extended inversion problem' which is solved by theta as well. Such
cases of degenerations were considered also by Baker (1897) and even
by Abel (see his \OE{}uvres 1881, {\bf I}: p.\,170--\ldots{}).

\subsection{On the $\Theta$-series}
\noindent
It is to be noticed, that contrary to commonly accepted (?) viewpoint,
the general $\Theta$-series \reff{B1}, as well as its argument
$\boldsymbol{z}=\widetilde{\boldsymbol{\omega}}(\mathcal{P})-\boldsymbol{e}$,
ought to be considered not as a special function or
formal generalization of Jacobi's $\theta$-function
but {\em regularly\/} derivable fundamental object for
explicit representation of all Abelian integrals, meromorphic/uniformizing functions, exponential Baker--Akhiezer-functions, and the theory as a whole%
\footnote{Jacobi's case  $g=1$ is not exception (Tichomandritzky, M.
1884 \textit{Math. Ann.\/} {\bf XXV}, 197--202). Riemann and Baker
(1897) do not elucidate an origin of appearance the $\Theta$-series.
Hyperelliptic case, based on Weierstrassian lectures, was considered
in dissertation by Tikhomandritskii (1885).}. Main ideology belongs
to K.\,Weierstrass\ (1902: pp.\,513--538) and was expounded in the
books by Tikhomandritskii (1885; 1895: pp.\,199--232) before
publication of Weierstrassian lectures on Abelian transcendents, but
with explicit use of the Riemann surface. Primarily and naturally
arising non-canonical form of the $\Theta$-series is derived in
these works with basic properties and identities. The canonical
formula \reff{B1}, is obtained after the normalization \reff{B}. See
also the book by Clebsch \& Gordan (1866:  pp.\,193--198). For lack
of space, we do not pursue these important points here. This will be
written up elsewhere. By the same reason we have somewhat reduced
exposition between formulas \reff{canonical} and \reff{inter}.

\subsection{Algebraically integrable Hamiltonian systems}

\noindent
In what concerns an inverse transition in the thesis on p.\,\pageref{thesis},
i.\,e. transition from an algebraically  integrable Hamiltonian
system
to some spectral problem, the answer is positive though  somewhat noneffective. We assume  that separability variables $\{\gamma,\,\mu\}$
sitting on a  curve, say  \reff{dubrovin}, exist.
We can construct the Baker--Akhiezer
function \reff{PsiTheta} and the Schr\"odinger
operator (it certainly exists) for it.
The computational part of the Liouville theorem is based on the Hamilton--Jacobi
theory of canonical transformations.
The variables $\gamma$'s are  always taken as poles of the $\Psi$:
$$
\gamma_k^{}(x):\qquad
\Theta\big(\widetilde{\boldsymbol{\omega}}(\infty)+
x\boldsymbol{U}-\boldsymbol{D}\big)=0
$$
since they solve the problem \reff{jacobi}, \reff{canonical}.
Ostrogradskii's variables $\{u,u_x,\ldots\}$ are constructed
explicitly by the trace formulas. Isospectral $t$-deformations of
finite-gap potentials are readily included in \reff{omega}: the
quantities $d_k$ become linear functions of
$x\boldsymbol{U}+t\boldsymbol{V}$. Accordingly, there is not
essential difference between $x$- and $t$-equations for the $\Psi$.
Both of these equations (Lax' pair) may be thought of as quadrature
integrable (factorizable) spectral problems/pencils:
Ermakov--Drach's constant $\mu$ becomes an eigenvalue of the second
commuting operator connected through the curve with the first one:
$$
\widehat{\!\boldsymbol{A}}(\pot;\partial_x)\Psi(x;\lambda)=
\mu(\lambda)\Psi(x;\lambda).
$$
There are infinitely many Abelian functions built from
$\{\gamma,\mu\}$ (followed by physical coordinates
$\{p(\gamma,\mu),\,q(\gamma,\mu)\}$) and expressible through the
$\Theta$'s. Non-constructiveness can appear because transformations
between various dynamical systems are not necessary to be canonical.
However canonicity is not a necessary attribute of quadrature
integrability. {\em One\/} fundamental Abelian function  determines
all the transformations---the potential $u$.

\begin{acknowledgements}
The author thanks Professor J.\,C.\,Eilbeck and Professor
E.\,Previato for numerous discussions and  EP for hospitality in
Boston University where the work was carried out.

The project was supported by  NSF/NATO-grant DGE--0209549.
\end{acknowledgements}

\appendix{Note on a transformation of Riccati's equation}

\centerline{R.\,Rawson. \textit{The Messenger of Mathematics}
{\bf XII} (1883), 34--36.}
\bigskip

\quad\; Riccati's equation

\medskip

\hfill
$
\displaystyle
\frac{dy_1^{}}{dx_1^{}}+a\,{y_1^{}}^2=\phi(x_1^{})
\ldots\ldots\ldots\ldots\ldots
\ldots\ldots\ldots\ldots\;(1)
$

\smallskip

\noindent
is readily transformed into

\smallskip

\hfill
$
\displaystyle
\frac{dy}{dx}+a\,y^2=\frac{a_1^{}}{a}\,\phi(P)\left(\frac{dP}{dx}
\right)^{\!2}
+\frac{3}{4\,a}\!\left(\frac{\,\mfrac{d^2P}{dx^2}\,}{\mfrac{dP}{dx}}\right)^
{\!\!2}
-\frac{1}{2\,a}\frac{\,\mfrac{d^3P}{dx^3}\,}{\mfrac{dP}{dx}}
\ldots\ldots\ldots\;(2)
$

\smallskip

\noindent
by means of the two equations

\smallskip

\hfill
$
\displaystyle
x_1^{}=P,\quad \mbox{a function of $x$}
\ldots\ldots\ldots\ldots\ldots
\ldots\ldots\ldots\;(3),
$

\bigskip

\hfill
$
\displaystyle
2\,a_1^{}\!\left(\frac{dP}{dx}\right)^{\!2}=2\,a\,y\,\frac{dP}{dx}+
\frac{d^2P}{dx^2}
\ldots\ldots\ldots\ldots
\ldots\ldots\ldots\;(4).
$

\smallskip

If, therefore, either (1) or (2) can be integrated, then the other can be
integrated also by means of (3) and (4).

An interesting case of the above transformation is when

\smallskip

\hfill
$
\displaystyle
\phi(x_1^{})=b\,{x_1^{}}^n+\frac{c}{{x_1^{}}^2}
\,.\ldots\ldots\ldots\ldots\ldots
\ldots\ldots\ldots\;(5),
$

\bigskip

\hfill
$
\displaystyle
P=x^m\ldots\ldots\ldots\ldots\ldots
\ldots\ldots\ldots\ldots\ldots\ldots\;(6).
$

We then have

\hfill
$\displaystyle
\frac{dy_1^{}}{dx_1^{}}+a_1^{}{y_1^{}}^2=b\,{x_1^{}}^n+\frac{c}{{x_1^{}}^2}
\;\ldots\ldots\ldots\ldots\ldots\ldots\ldots\ldots\;(7),
$

\bigskip

\hfill
$\displaystyle
\frac{dy}{dx}+a\,y^2=\frac{a_1^{}b\,m^2}{a}\,x^{(n+2)m-2}+
\frac{(4\,a_1^{}c+1)\,m^2-1}{4\,a\,x^2}
\;\ldots\ldots\ldots\ldots\;(8),
$
$$
x_1^{}=x^m,
$$
$$
2\,a_1^{}m\,x^{m-1}\,y_1^{}=a\,y+\frac{m-1}{2\,x}.
$$

In equation (7) let $c=0$, and $n=-\mfrac{4\,p}{2\,p \pm 1}$, where $p$ is an
integer, then (7) and (8) become

\smallskip

\hfill
$
\displaystyle
\frac{dy_1^{}}{dx_1^{}}+a_1^{}{y_1^{}}^2=b\,x_1^{-\frac{4p}{2p\pm1}}
\ldots\ldots\ldots\ldots\ldots\ldots\ldots\ldots\;(9),
$

\smallskip

\hfill
$
\displaystyle
\frac{dy}{dx}+a\,y^2=\frac{a_1^{} b\,m^2}{a}\,x^{\frac{\pm 2m}{2p\pm 1}-2}
+\frac{m^2-1}{4\,a\,x^2}
\ldots\ldots\ldots\ldots\ldots\;(10);
$

\smallskip

\noindent
and therefore since (9) is soluble, so also is (10).

Equation (10) is made linear by putting
$$
y=\frac{1}{a}\frac{dz}{z\,dx},
$$
and we thus find
$$
\frac{d^2z}{dx^2}=\left\{a_1^{} b\,m^2\,x^{\frac{\pm 2m}{2p\pm1}-2}
+\frac{m^2-1}{4\,x^2} \right\}z,
$$
or

\smallskip

\hfill
$\displaystyle
\frac{d^2z}{dx^2}=\left\{\phantom{\frac{{}^{\ds\mathstrut}}{\ds\mathstrut}}
 \right. \!\!\!\!a_1^{} b\,(2\,q+1)^2\,x^{\frac{\pm 2(2q+1)}{2p\pm 1}-2}
+\frac{q\,(q+1)}{x^2} \!\!\!\left.\phantom{\frac{{}^{\ds\mathstrut}}{\ds\mathstrut}}\right\}z\;
\ldots\ldots\ldots\;(11),
$

\smallskip

\noindent
where
$$
m=2\,q+1.
$$

The equations of transformation then become
$$
x_1^{}=x^{2q+1},
$$
$$
2\,a_1^{}(2\,q+1)\,x^{2q}\,y_1^{}=a\,y+\frac{q}{x}.
$$

The differential equation (11) is of some interest as it includes as a
particular case the well-known differential equation
$$
\frac{d^2z}{dx^2}=\left\{a^2  + \frac{q\,(q+1)}{x^2}\right\}z.
$$

\hfill {\sc Robert Rawson.}

\appendix{Notation to \S\,7}

The $\Theta$-function corresponds to the symmetrical matrix $B_{jk}$ \reff{B}
\begin{equation}\label{B1}
\Theta(\boldsymbol{z}|B)\equiv\Theta(z_1^{},\ldots\!, z_g|B)=
\sum\limits_{\mbox{\footnotesize\raisebox{-0.11em}{$\boldsymbol{N}$}}
\in
\mbox{\footnotesize\,\raisebox{-0.11em}{$\mathbb{Z}$}}^g}
\!\re^{\pi \ri\, \langle B\boldsymbol{N},\boldsymbol{N}\rangle +
2\pi \ri\,\langle\boldsymbol{N},\boldsymbol{z}\rangle},
\end{equation}
where
$\langle B\boldsymbol{N},\boldsymbol{N}\rangle=
\sum\ds B_{jk}N_j N_k$ and
$\langle\boldsymbol{N},\boldsymbol{z}\rangle=N_1^{}z_1^{}+\cdots+N_g z_g$.
Canonical base of cycles $(a_j^{},\,b_j^{})$
on $\boldsymbol{\mathfrak{S}}$ is chosen according to the intersection
scheme:
$a_j^{} \circ b_k^{}=\delta_{jk}$.
The normal Abelian integrals of the 2, 3-rd kind and their periods have the form:
$$
\widetilde\Omega=\frac{1}{\tau-\tau_{\sinfty}}+
c_1^{}(\tau-\tau_{\sinfty})+\cdots,\qquad
\Oint\limits_{a_k^{}}\rd\widetilde\Omega=0,\qquad
\Oint\limits_{b_k^{}}\rd\widetilde\Omega=2\,\pi\,\ri\,U_k,
$$
$$
\ds\widetilde\Pi_{\alpha,\nu}(\tau)=
\ln\mfrac{\tau-\tau_{\alpha}}{\tau-\tau_{\nu}}
+\cdots,\qquad
\Oint\limits_{a_k^{}}\rd\widetilde\Pi_{\alpha,\nu}=0,\qquad
\Oint\limits_{b_k^{}}\rd\widetilde\Pi_{\alpha,\nu}=2\,\pi\,\ri\,
\big(\widetilde\omega_k^{}(\nu)-\widetilde\omega_k^{}(\alpha)\big).
$$

\label{lastpage}
\end{document}